\documentclass[11pt,ams]{article}%
\usepackage{geometry}
\usepackage{dsfont}
\usepackage{amsmath}
\usepackage{amsfonts}
\usepackage{amssymb}
\usepackage{graphicx}%
\usepackage{subfigure}
\usepackage{float,pslatex,braket}
\usepackage[section]{placeins}
\usepackage{verbatim}
\usepackage{hyperref}

\newcommand{\p}{\partial}

\begin{document}

\date{}
\title{\textbf{On the transition temperature(s) of magnetized two-flavour holographic QCD}}
\author{\textbf{N.~Callebaut}\thanks{ncalleba.callebaut@ugent.be}\,\,,
\textbf{D.~Dudal}\thanks{david.dudal@ugent.be}\,\,\\[2mm]
{\small \textnormal{\it Ghent University, Department of Physics and Astronomy, Krijgslaan 281-S9, 9000 Gent, Belgium}\normalsize}}
\maketitle
\begin{abstract}
\noindent During heavy ion collisions, high temperatures and strong magnetic fields are generated. We employ the gauge-gravity duality to study the $N_f=2$ QCD phase diagram under these extreme conditions in the quenched approximation, in particular we use the non-antipodal Sakai-Sugimoto model (SSM). We take the different coupling of up and down flavours to the magnetic field into account geometrically, resulting in a split of the chiral phase transition according to flavour. We discuss the influence of the magnetic field on the chiral temperatures ---in physical GeV units--- in terms of the choice of the confinement  scale in the model, extending hereby our elsewhere presented discussion of fixing the non-antipodal SSM parameters to the deconfinement phase.
The flavour-dependent $(T,L,eB)$ phase diagram, with variable asymptotic brane-antibrane separation $L$, is also presented, as a direct generalization of the known $(T,L)$ phase diagram of the non-antipodal SSM at zero magnetic field. In particular, for sufficiently small $L$ we are probing a NJL-like boundary field theory in which case we do find results very reminiscent of the predictions in NJL models.
\end{abstract}

\baselineskip=13pt

\section{Motivation}
It is already known for some time that gigantic magnetic fields occurred during the cosmological electroweak phase transition \cite{Vachaspati:1991nm}, but recently, it has also become clear that magnetic fields up to $10^{15}$ Tesla can occur shortly during relativistic heavy ion collisions, see e.g.~\cite{Kharzeev:2007jp,Skokov:2009qp}. To put such a field strength in the correct perspective, a magnetar, or highly magnetized neutron star, reaches ``merely'' $10^{9}$ Tesla. The interest in QCD studies under these extreme conditions has therefore increased considerably. As strong coupling effects are relevant in the setting of interest, we must rely on nonperturbative tools to study the relatively new realm of QCD physics in a strong magnetic background. This has initiated a vast amount of original research, let us refer to \cite{Kharzeev:2012ph} and references therein for a recent review. In particular, confinement and chiral symmetry breaking are two typical nonperturbative QCD effects that can be affected. It is common knowledge that QCD deconfines at a certain temperature $T_c$, while chiral symmetry is restored at $T_\chi$. We shall be concerned with QCD in the chiral limit here, i.e.~we ignore the bare quark masses, in which case a clear-cut chiral transition exists. In real life QCD, with massive dynamical flavours, only approximate order parameters can be defined for both chiral restoration ($\sim$ chiral condensates) and deconfinement ($\sim$ Polyakov loop), leading to cross-over behaviour rather than sharp phase transitions. Once a hot debate whether $T_c$ and $T_\chi$ coincided or not, see \cite{Aoki:2006we,Bazavov:2009zn} for 2 views on the $N_f=2+1$ case, it is by now accepted they are close in the cross-over region. These results were obtained using lattice simulations, a powerful ally to access the nonperturbative QCD sector.

More recently, a vivid discussion evolved around the possibility that $T_c$ and $T_\chi$ separate under the influence of a constant magnetic background field, $\mathbf{B}=B\mathbf{e}_z$, and this for $N_f=2$ QCD, see e.g.~\cite{Mizher:2010zb,Gatto:2010pt,D'Elia:2010nq}. Although the $N_f=2$ lattice results of \cite{D'Elia:2010nq} indicated a weak rise in the transition temperatures $T_c$ and $T_\chi$, both remained compatible with each other (a split of $\sim 2\%$), while the various analytical model based results were inconclusive on the matter, as different results were obtained per QCD model  \cite{Mizher:2010zb,Gatto:2010pt}. Somewhat later, a more thorough lattice study appeared using $N_f=2+1$ flavours with physical masses, leading to a much more complicated behaviour in the chiral/deconfinement (pseudo-)order parameters and ensuing critical temperatures \cite{Bali:2011qj}. It was motivated that the reported behaviour ---where contrasting with the results of \cite{D'Elia:2010nq}--- should be traced back to the lighter dynamical flavours and partially also to the present strange flavour, as the up ($u$) and down ($d$) quark of \cite{D'Elia:2010nq} were considerably heavier. Soon after, the first analytical papers appeared trying to explain the state-of-the art lattice data using backreacting pion dynamics \cite{Fraga:2012fs}. The naive reason for expecting a split between $T_c$ and $T_\chi$ was the expected enhancement of chiral symmetry breaking due to a magnetic field ---the so-called chiral magnetic catalysis \cite{Miransky:2002rp}. The results of \cite{Bali:2011qj} showed a more subtle picture:
the magnetic catalysis was confirmed for temperatures (sufficiently) below $T_c$, but for larger temperatures the (averaged over up and down) chiral condensate displayed a non-monotonous shape, a feature translated into a similar behaviour in the transition temperature. This observation of an ``inverse magnetic catalysis'' seems to depend crucially on taking into account quark backreaction effects, see also \cite{Bruckmann:2013oba}, so we do not expect it to appear in the unquenched Sakai-Sugimoto setting we will use.

Since a magnetic field couples to the up and down flavours with another strength, as they carry different electric charges, it seems natural that the up and down chiral restoration temperatures can be different, as well as the magnetic catalysis itself.  We recall that the classical chiral structure of QCD with and without magnetic field is different, since coupling a magnetic field to the quarks reduces $U(2)_L\times U(2)_R$ to $\left[U(1)_L\times U(1)_R\right]^u\times \left[U(1)_L\times U(1)_R\right]^d$, so that the eventually broken chiral invariances $U(1)_A^u$ and $U(1)_A^d$ can experience a different restoration temperature. Lattice simulations indeed confirm a larger value for the $\braket{\overline u u}$ than for the $\braket{\overline d d}$ chiral condensate at $T=0$ \cite{D'Elia:2011zu}, as does the $N_f=2$ Nambu-Jona-Lasinio (NJL) model \cite{Boomsma:2009yk}. It would appear natural that $T_\chi^{u}$ should consequently be larger than $T_\chi^{d}$. The splitting of degenerate order parameters, like $\braket{\overline u u}$ and $\braket{\overline d d}$ at $eB=0$, when an external field is switched on, is not that unfamiliar. In certain exotic superconductors, e.g.~$\text{Sr}_2\text{RuO}_4$, a similar phenomenon occurs \cite{Mackenzie:2003zz}.

Here, we will use the gauge-gravity duality, a powerful tool to analytically study certain aspects of strongly coupled gauge theories \cite{Maldacena:1997re}, to shed further light on the possible $(T,eB)$ QCD phase diagram. In particular, we rely on the non-antipodal Sakai-Sugimoto model (SSM) \cite{Sakai:2004cn} with 2 quenched flavours, represented by two D8-$\overline{\text{D8}}$ probe brane-antibrane pairs, in the chiral limit and with 3 colours. Going beyond the quenched approximation and/or including massive dynamical flavours leads to utter complications \cite{Burrington:2007qd,Dhar}. We take into due account the different coupling of up and down flavour to the magnetic field, leading to a different probe brane geometry per flavour. We compare our findings with the lattice results of \cite{Ilgenfritz:2012fw}. Although those results are referring to two-colour QCD, it contains an extrapolation to the chiral limit, which is the closest the available lattice results come to the $N_f=2$ SSM\footnote{We will use $N_c=3$ colours to get explicit numbers, where in principle the limit $N_c\to\infty$ is always understood at the holographic level. Setting $N_c=2$ or $N_c=3$ will not induce any qualitative change in the SSM results.}. In \cite{Ilgenfritz:2012fw}, no manifest split between $T_c$ and $T_\chi$ was reported, whilst the chiral condensate increases monotonically with the applied magnetic field for all temperatures in the confinement phase. We do present similar results here using prefixed values for the string theory parameters of the SSM. With prefixed, we mean that a few physical QCD input values at zero magnetic field are chosen to match the corresponding SSM predictions. These results are a generalization to the 2 flavour case of the single flavour analysis of \cite{Johnson:2008vna}. To extend the scope of our analysis, we will also allow that the asymptotic D8-$\overline{\text{D8}}$ separation $L$ ---or equivalently, the confinement scale $M$---  can vary and as such we construct the magnetic generalization of the $(T,L)$ phase diagram of Figure 7 in the original work \cite{Aharony:2006da} concerning the SSM phase diagram. All results are presented in GeV units to make comparison with other QCD approaches more direct.

\section{Setup}

\subsection{The Sakai-Sugimoto model}
\begin{figure}[h]
  \centering
  \scalebox{0.7}{
  \includegraphics{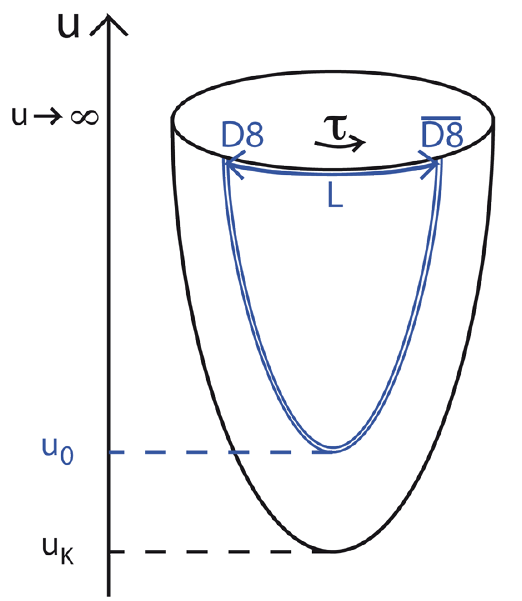}}
  \caption{The Sakai-Sugimoto model.}\label{SS}
\end{figure}

At zero temperature, the SSM \cite{Sakai:2004cn} involves a system of $N_f$ pairs of $\text{D}8$-$\overline{\text{D8}}$ flavour probe branes placed in the D4-brane background
\begin{eqnarray}\label{D4}
ds^2 &=& \left(\frac{u}{R}\right)^{3/2} (\eta_{\mu\nu}dx^\mu dx^\nu + f(u)d\tau^2)+ \left(\frac{R}{u}\right)^{3/2}
\left( \frac{du^2}{f(u)} + u^2 d\Omega_4^2 \right),\qquad f(u)=1-\frac{u_K^3}{u^3} \,,
\end{eqnarray}
where $R^3 = \pi g_s N_c \ell_s^3$, with $g_s$ ($\ell_s$) the string constant (length). There is a natural cut-off at $u = u_K$, which ensures confinement in the dual field theory living at the boundary $u\to\infty$. A smooth cut-off is realized if $\tau$ has a periodicity $\delta \tau = \frac{4\pi}{3} R^{3/2}u_K^{-1/2} = 2\pi M^{-1}$ with $M$ the compactification scale. A $U(N_f)_L \times U(N_f)_R$ gauge theory resides on the stack of coinciding $\text{D}8$-$\overline{\text{D8}}$ flavour pairs, which corresponds to the global chiral symmetry in the dual QCD-like theory. The cigar-shape of the ($u,\tau$)-subspace of the D4-brane background forces the embedding of the flavour branes to be $\cup$-shaped, which signals the breaking of chiral symmetry $U(N_f)_L \times U(N_f)_R \rightarrow U(N_f)_V$ as the merging of the D8-branes and $\overline{\text{D8}}$-branes at the value $u = u_0\geq u_K$. The value $u_0$ is directly related to the asymptotic separation $L$ (at $u\rightarrow \infty$) between D8- and $\overline{\text{D8}}$-branes, indicated in Figure \ref{SS}. For values of $u_0$ greater than or equal to $u_K$ the embedding is respectively non-antipodal or antipodal.

\begin{figure}[h]
\centering
 \subfigure[]{
\includegraphics[scale=0.7]{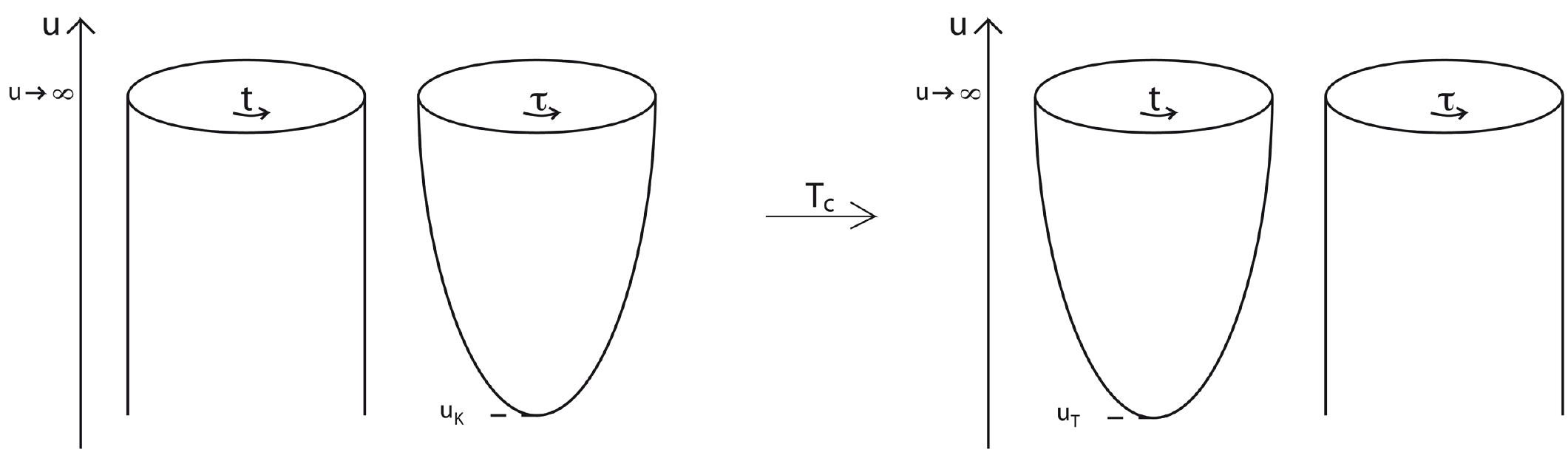}
}
\subfigure[]{
\includegraphics[scale=0.7]{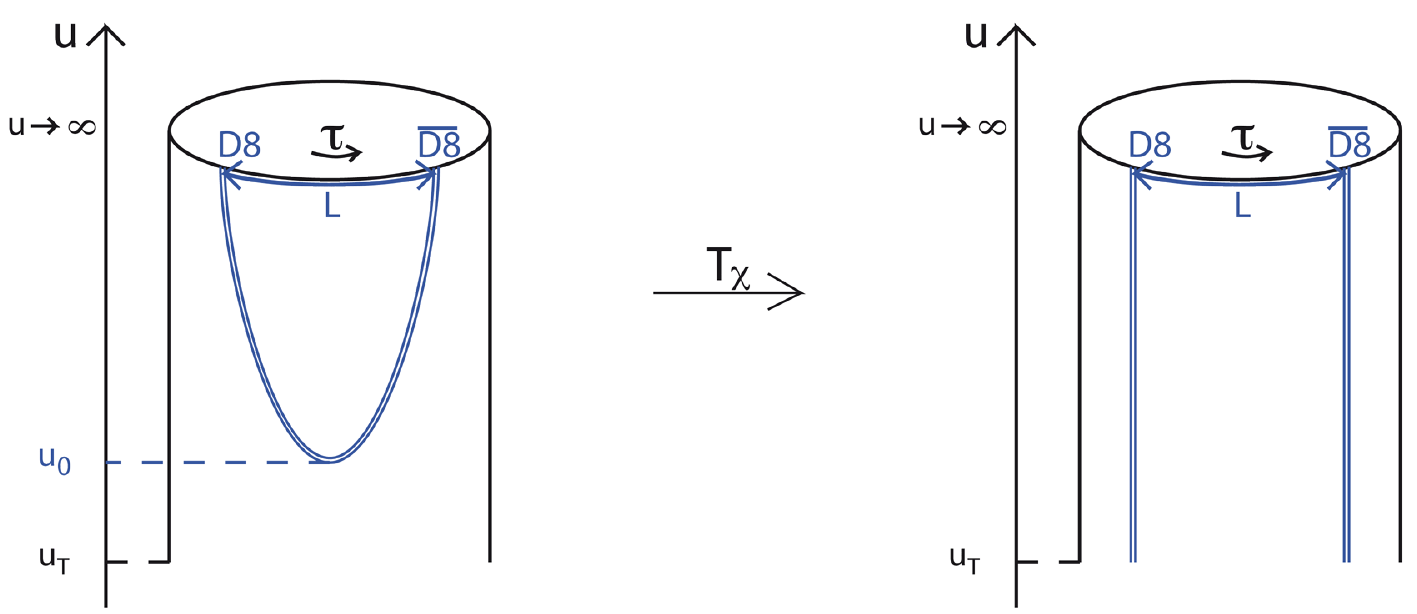}
}
\caption[]{(a) Deconfinement transition at $T_c$ and (b) chiral symmetry restoration at $T_\chi (\geq T_c)$.} \label{figure2}
\end{figure}

At finite temperature there are two regular Euclidean backgrounds with the same asymptotic geometry that compete with each other in the partition function, the Wick-rotated version of the D4-brane background \eqref{D4} and a black D4-brane background \cite{Aharony:2006da}. The Wick-rotated D4-brane background has a cigar-shaped $(u,\tau)$-subspace and a cilinder-shaped $(u,t)$-subspace, with the periodicity of the $t$-circle arbitrary and equal to the inverse temperature, $\beta = T^{-1}$, and the periodicity of the $\tau$-circle fixed to $\delta \tau = \frac{4\pi}{3} R^{3/2}u_K^{-1/2}$, while the black D4-brane background
 \begin{eqnarray}
ds^2 &=& \left(\frac{u}{R}\right)^{3/2} (\hat f(u) dt^2 + \delta_{ij}dx^i dx^j + d\tau^2) +\left(\frac{R}{u}\right)^{3/2}
\left( \frac{du^2}{\hat f(u)} + u^2 d\Omega_4^2 \right),\qquad  \hat f(u) = 1-\frac{u_T^3}{u^3}\,,
\end{eqnarray}
has a cigar-shaped $(u,t)$-subspace and a cilinder-shaped $(u,\tau)$-subspace, now with the periodicity of the $\tau$-circle arbitrary, but the periodicity of the $t$-circle fixed to $
\delta t = \beta = T^{-1}= \frac{4\pi}{3} R^{3/2} u_T^{-1/2} $. These two backgrounds are identical, mo\-du\-lo a redefinition of coordinates $t$ and $\tau$, when $\delta \tau$ equals $\beta$, which happens at the deconfinement transition temperature $T_c = \frac{3}{4\pi} R^{-3/2}  u_K^{1/2}$.
In the deconfining phase\footnote{
Let us remark here that in \cite{Mandal:2011ws} some problems concerning the identification of the deconfined phase with the black D4-brane background were discussed, and instead a different background was proposed, namely a localized D3-soliton geometry. 
To make calculations of the critical temperatures feasible, it is however necessary to consider a high-temperature approximation of that background. There are moreover some subtleties concerning the inclusion of flavour branes, but the end result for the $(T,L)$ phase diagram, see Fig. 10 of \cite{Mandal:2011ws}, gives a qualitatively similar result as the `old' SSM. Based on this observation, we can expect that qualitative features of the $eB$-dependence of the chiral transition temperatures are not unlikely to be similar in both backgrounds. It would be interesting to check this explicitly, but for the reasons mentioned above we will consider the simpler black D4-brane background in this paper. We thank Takeshi Morita for discussion on this point. }
 the embedding of the flavour branes is no longer forced to be $\cup$-shaped, as the $(u,\tau)$-space is no longer cigar-shaped. At a certain value of the temperature, $T_\chi \geq T_c$, it will become energetically favourable for the flavour branes to fall straight down instead of merging in a $\cup$-shape, indicating chiral symmetry restoration (see Figure~\ref{figure2}). The essential features of QCD, chiral symmetry breaking and confinement, plus the ensuing chiral restoration and deconfinement at sufficiently large temperatures are thus nicely resembled by the SSM, amongst other QCD phenomenology \cite{Sakai:2004cn}. To make further explicit contact with QCD, we determine the string related parameters of the SSM in physical (GeV) units, so that an explicit comparison with other approaches comes within range.

\subsection{Fixing holographic parameters at $eB=0$}
In \cite{Sakai:2004cn} the independent parameters $M$ and $\kappa$ of the antipodal ($u_0=u_K$) SSM were fixed to $M \approx 0.949$ GeV and $\kappa = \lambda/(72 \pi^3) \approx 0.00745$, with $\lambda = g^2_{YM} N_c$ the 't Hooft coupling, by matching to the QCD input values
\begin{eqnarray} \label{fixing}
\quad f_\pi = 0.093  \mbox{ GeV} \quad \mbox{and} \quad m_\rho = 0.776 \mbox{ GeV}
\end{eqnarray}
for the pion decay constant $f_\pi$ and the $\rho$ meson mass $m_\rho$.
Due to the relations $u_K=M^{-1}$, $R^3 = \frac{9}{4} M^{-3}$ and $g_s^{-1} \ell_s^{-3} = \frac{4\pi}{3} M^3$ \cite{Sakai:2004cn} all other parameters of the model are then fixed as well. Note that throughout this paper we set $N_c = 3.$

\begin{figure}[h]
\centering
 \subfigure[]{
\includegraphics[scale=0.65]{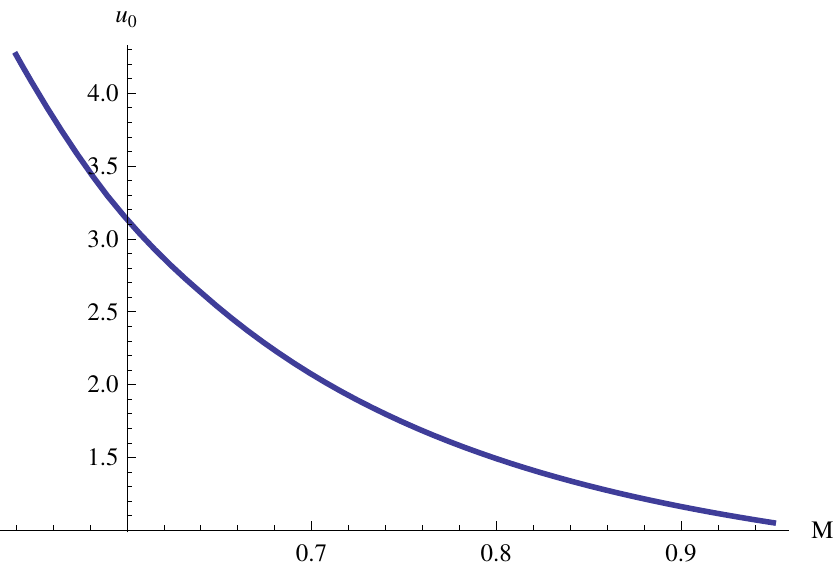} \label{a}
}
\subfigure[]{
\includegraphics[scale=0.65]{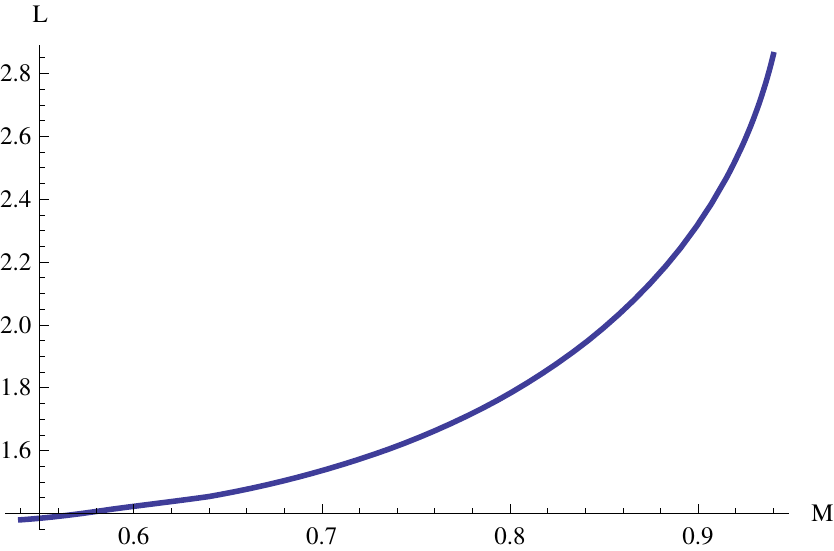} \label{b}
}
 \subfigure[]{
\includegraphics[scale=0.65]{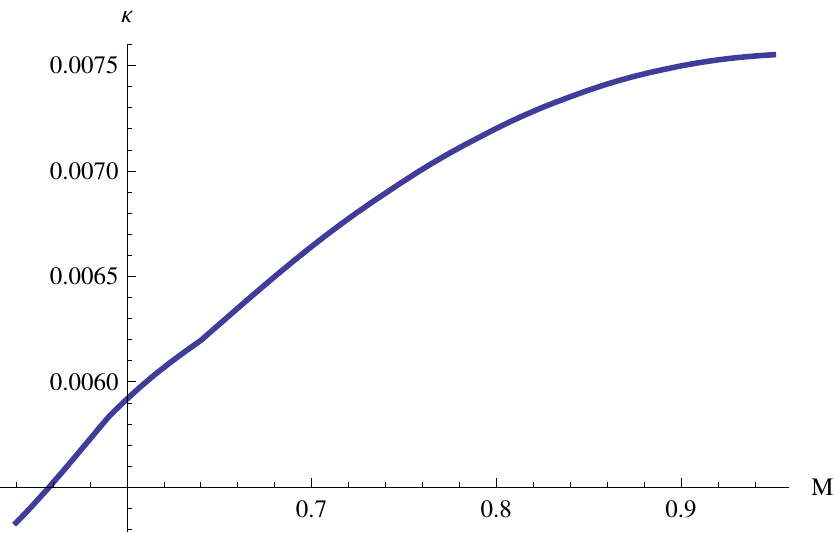} \label{c}
}
\caption[]{(a) Values of $u_0$ ($\text{GeV}^{-1}$) and (b) corresponding values of $L$ ($\text{GeV}^{-1}$)  compatible with $m_\rho=0.776$ GeV. (c) Values of $\kappa$ compatible with $m_\rho=0.776$ GeV and $f_\pi = 0.093$ GeV.}  \label{parameters}
\end{figure}

In the non-antipodal case ($u_0>u_K$) the matching conditions (\ref{fixing}) do not fix all freedom, since there is one extra parameter present, $u_0$.
From the eigenvalue equation which determines $m_\rho$ holographically,
\begin{equation} \label{}
\partial_z \left( \frac{3}{u_0} u_z^{1/2} \gamma'^{-1} \partial_z \psi_\rho \right) = - \frac{4}{3}u_0 u_z^{3/2} \gamma' R^3 m_\rho^2 \psi_\rho, \quad \psi_\rho'(0)=0, \quad \psi_\rho(\pm \infty) = 0,
\end{equation}
with
\begin{eqnarray}
\gamma'(z) = \sqrt{\frac{z^2}{u_z^{5}(u_z^3 - u_K^3) - (u_0^8 - u_0^5 u_K^3)}}, \quad u_z^3 =  u_0^3 + u_0 z^2,
\end{eqnarray}
we extract the values of $u_0$ that, for a given $M$, lead to $m_\rho = 0.776 \mbox{ GeV}$. The resulting function $u_0(M)$ is plotted in Figure~\ref{parameters} for a range of $M$  ---the maximum value of $M$ corresponding to the limiting case $u_0 \rightarrow u_K = 1/M$--- alongside with the function $L(M)$ for the corresponding asymptotic separation between branes and anti-branes, determined from
\begin{eqnarray}
L    =  \int_{u_0}^\infty du  \left(\frac{R}{u}\right)^{3/2} f^{-1} \sqrt{\frac{u_0^8 f_0}{u^8 f - u_0^8 f_0}},
\end{eqnarray}
with $f(u_0)$ denoted as $f_0$. Next, demanding that the SSM-prediction for the pion decay constant \cite{Callebaut:2011ab},
\begin{eqnarray}
f_\pi(M,u_0,\kappa) = \sqrt{ \frac{4}{3} \kappa M^{7/2} \frac{3}{u_0}\left( 2 \int_{0}^\infty dz \frac{\gamma'}{(u_0^3+u_0 z^2)^{1/6}} \right)^{-1} },
\end{eqnarray}
equals $0.093$ GeV, leads to the function $\kappa(M)$ of allowed values for $\kappa$, as plotted in Figure~\ref{c}.
The string tension $(2\pi\alpha')^{-1} = 8\pi^2 M^2 \kappa(M)$ is then also known as a function of $M$.

\begin{figure}[h]
\centering
\subfigure[]{
\includegraphics[scale=0.65]{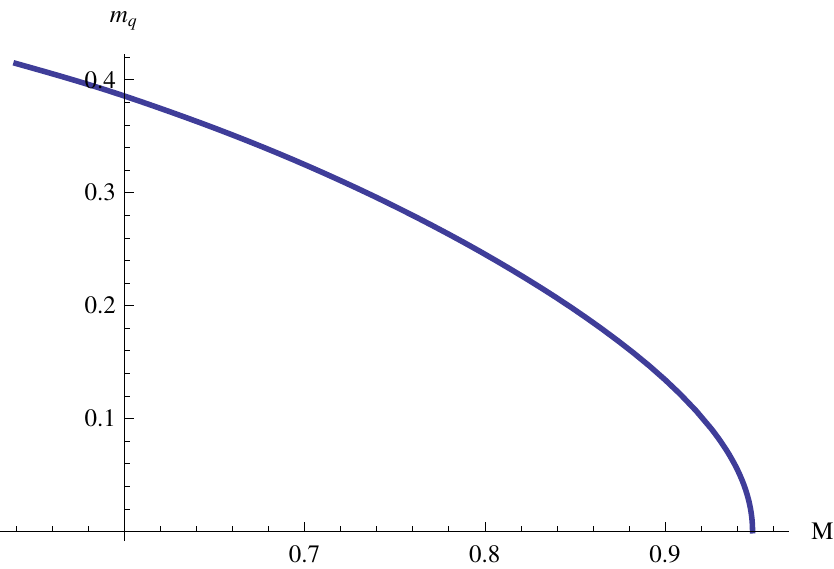}
}
\caption[]{SSM-prediction for $m_q$ ($\text{GeV}$) as a function of the confinement scale $M$ ($\text{GeV}$),  compatible with $m_\rho=0.776$ GeV and $f_\pi = 0.093$ GeV.} \label{mqMK}
\end{figure}

The remaining freedom of choosing the mass scale $M$ can be fixed for example by matching the SSM-prediction for the constituent quark mass,
\begin{equation}  \label{mq}
m_q(M,u_0,\kappa) =8\pi^2 M^2 \kappa \int_{1/M}^{u_0} du \frac{1}{\sqrt{1 - \frac{1}{(Mu)^3}}},
\end{equation}
to a phenomenologically reasonable value, as can be read off from Figure~\ref{mqMK}.

In \cite{Callebaut:2011ab} we opted to reproduce $m_q = 0.310$ GeV, leading to the following set of fixed holographic parameters:
\begin{equation} \label{values}
M \approx 0.7209 \mbox{  GeV}, \quad L \approx  1.574 \mbox{  GeV}^{-1} \quad
\mbox{and}\quad \kappa  = \frac{\lambda N_c}{216 \pi^3} \approx 0.006778.
\end{equation}
With these values, the effective QCD string tension is computed to be $\sigma \approx 0.19 \text{ GeV}^2$ \cite{Callebaut:2011ab}, in excellent agreement with the standard lattice value, $\sigma \approx 0.18$-$0.19 \text{ GeV}^2$, extracted from \cite{Bali:1992ru}. Said otherwise, we could equally well have selected the string tension as our third QCD input value instead of the constituent quark mass.

Here, we will however opt to leave $M$ variable, or equivalently $L$ via Figure~\ref{b}, with the eye on drawing the $(T,L,eB)$ phase diagram later, and, more importantly, with the idea that the choice of $M$ or $L$ should be left free, as it determines the choice of holographic theory: $L$ very small ($\sim$ $\delta \tau=\frac{2\pi}{M}$ large $\sim M$ small) corresponding to an NJL-type boundary field theory  \cite{Antonyan:2006vw,Preis:2010cq,Aharony:2006da} versus $L=\delta \tau/2$ maximal ($\sim M$ maximal) corresponding to a maximal probing of the gluon background, i.e.~the original antipodal SSM.

\FloatBarrier

\subsection{Applying a magnetic field}
The next step is to introduce a magnetic background. To this purpose, we take a closer look at the $U(N_f)$ gauge field $A_m(x^\mu,u)$ ($m=0,1,2,3,u)$ living on the D8-brane, with the non-Abelian DBI-action\footnote{As the 't Hooft
coupling $\lambda$ is large, we ignore
the Chern-Simons part of the action in the analysis, being a factor
$1/\lambda$ smaller than the DBI-action.} \cite{Tseytlin:1997csa}
\begin{eqnarray}\label{DBI}
S &=& -2T_8 \int d^4x du \hspace{1mm}  \int \epsilon_4 \hspace{1mm} e^{-\phi}\hspace{1mm} \textrm{STr}
\sqrt{-\det \left[g_{mn}^{D8} + (2\pi\alpha') iF_{mn} \right]},
\end{eqnarray}
where the factor 2 in front makes sure that we integrate (in $u$) over both halves of the $\cup$-shaped D8-branes, $e^{\phi}=g_s (u/R)^{3/4}$, $\epsilon_4$ is the volume form of the unit 4-sphere in the background, $T_8 = 1/((2\pi)^8 \ell_s^9)$ the D8-brane tension,  $\text{STr}$ the
symmetrized trace defined as $\text{STr}(F_1 \cdots F_n) = \frac{1}{n!} \text{Tr}(\text{permutations of }F_1 \cdots F_n)$, $g_{mn}^{D8}$ the induced metric on the D8-branes,  $\alpha'= \ell_s^2$ the string tension, and $F_{mn}$ the usual
field strength; we use anti-Hermitian generators. As explained in \cite{Sakai:2004cn,Callebaut:2011ab}, one can slightly gauge the global $U(N_f)_V$ symmetry in the boundary field theory, i.e.~make $g(u\to\infty)\in U(N_f)_V$ $x$-dependent, so that $g\p_\mu g^{-1}\equiv \overline A_\mu(u\to\infty)$ corresponds to adding a background $U(N_f)_V$ field. Working in the $A_u=0$ gauge, the flavour gauge field can then be expanded as \cite{Sakai:2004cn}
\begin{eqnarray}
A_\mu(x^\mu,u) &=& \overline A_\mu (x^\mu) + \text{rest}\,,
\end{eqnarray}
where ``rest" refers to pions and vector mesons in the boundary QCD theory, which are irrelevant for the current purposes.

An electromagnetic background field $A_\mu^{em}$ can be switched on through
\begin{equation}
\overline A_\mu = -i e \mathbf{Q}_{em} A_\mu^{em} = -i e\left(
                                                        \begin{array}{cc}
                                                          2/3 & 0 \\
                                                          0 & -1/3 \\
                                                        \end{array}
                                                      \right)
 A_\mu^{em} = -i e
\left( \frac{\textbf{1}_2}{6}  + \frac{\sigma_3}{2}  \right)  A_\mu^{em}\,,
\end{equation}
with $\mathbf{Q}_{em}$ the charge matrix for the up- and down-quarks. The choice $A_\mu^{em}=x_1B\delta_{\mu2}$, or $\overline A_2^3 = x_1 e B$, $\overline A_2^0 =
\overline A_2^3 / 3$, ensures the desired constant magnetic background $\mathbf{B}=B\mathbf{e}_3$. The corresponding field strength tensor reads
\begin{eqnarray}
\overline F_{12}&=& \partial_1 \overline A_2 =  -i \left(\begin{array}{cc} \frac{2}{3}eB  & 0 \\ 0 & -\frac{1}{3}eB \end{array} \right) \equiv -i \left(\begin{array}{cc} \overline F_u  & 0 \\ 0 & \overline F_d \end{array} \right)\,.
\end{eqnarray}

\section{Results}

\subsection{$eB$-dependent embedding in confinement phase}

For completeness, we briefly summarize here the discussion of the $eB$-dependent embedding of the $N_f=2$ flavour branes in the confinement phase, i.e. the determination of $u' = du/d\tau$ for each flavour, presented in more detail in \cite{Callebaut:2011ab}.

The induced metric on the flavour branes is given by
\begin{eqnarray}
ds^2_{D8} &=& g_{mn}^{D8} dx^m dx^n \quad(m,n = 0 \cdots 8) \nonumber \\
&=& \left(\frac{u}{R}\right)^{3/2} \eta_{\mu\nu}dx^\mu dx^\nu + \left( \left(\frac{R}{u}\right)^{3/2} \frac{1}{f(u)} +  \left(\frac{u}{R}\right)^{3/2} \frac{f(u)}{u'^2} \right) du^2 + \left(\frac{R}{u}\right)^{3/2} u^2 d\Omega_4^2
\end{eqnarray}
where the metric components in flavour space are assumed to be different, to allow for a different response of both flavour branes to the magnetic field:
\begin{eqnarray}
g^{D8}
&=& \left( \begin{array}{cc} g^{D8}_u & 0 \\ 0 & g^{D8}_d \end{array} \right),
\end{eqnarray}
with the $u$-coordinate appearing in $g^{D8}_u$, resp.~$g^{D8}_d$ following the up-, resp.~down-brane, thus $u\in\left[u_{0,u},\infty\right]$ or $u\in\left[u_{0,d},\infty\right]$.
Inserting this metric-ansatz into the DBI-action \eqref{DBI}, along with the $\overline A_\mu$ background, gives  $S^{conf} =S_{u} + S_{d}$ with
\begin{eqnarray}
S_\ell &=&  -T_8 \mathcal{V}_4 V_4 g_s^{-1}  \hspace{1mm} 2\int_{u_{0,\ell}}^\infty du \hspace{1mm} u^4 \sqrt{ \frac{1}{f} \left(\frac{u}{R}\right)^{-3} + \frac{f}{u'^2} } \sqrt{A_\ell}\,,\quad
A_\ell =  1 + (2\pi\alpha')^2 \overline F_\ell^2  \left(\frac{R}{u}\right)^{3},
\end{eqnarray}
where the index $\ell$ refers to the up or down flavour, $u' = du/d\tau$, $\mathcal{V}_{4}=\int d^4x$ and $V_4$ is the volume of the unit 4-sphere. The DBI-action for the two D8-branes thus reduces to the sum of two Abelian actions, which is the geometric translation of the explicit breaking of chiral symmetry $U(2)_A \stackrel{eB}{\rightarrow} U(1)_A^u \times U(1)_A^d$.

Suppressing the $\ell$-index for the moment, $u'$ can be determined from the conserved
``Hamiltonians''
\begin{eqnarray}
&& H = u' \frac{\delta \mathcal L^\tau}{\delta u'} - \mathcal L^\tau =  \frac{-u^4 f \sqrt{A}}{\sqrt{\frac{u'^2}{f} \left(\frac{u}{R}\right)^{-3} + f}}\,\,, \quad \partial_\tau H = 0\,, \quad \mbox{with } \quad \mathcal L^\tau
= u^4  \sqrt{\frac{u'^2}{f} \left(\frac{u}{R}\right)^{-3} + f} \hspace{1mm} \sqrt{A}\,
\end{eqnarray}
by assuming $\cup$-shaped embeddings, $u' = 0$ at $u=u_{0}$ (with $A(u_0)$ and $f(u_0)$ denoted as $A_0$ and $f_0$, for each flavour). In the confinement phase, the $eB$-dependent action on each flavour brane is then given by
\begin{eqnarray}\label{actionmagneticfield}
S_\ell &=&  -T_8 \mathcal{V}_4 V_4 g_s^{-1} \hspace{1mm} 2 \int_{u_{0,\ell}}^\infty du \hspace{1mm}  R^{3/2} u^{5/2} \sqrt{A_\ell} \sqrt{\gamma_{B,\ell}}\,,  \quad \gamma_{B,\ell}(u) = \frac{u^8 A_\ell}{u^8 f A_\ell - u_{0,\ell}^8 f_0 A_{0,\ell}},
\end{eqnarray}
and the asymptotic separations $L=2\int du/u'$ become (suppressing $\ell$ for notational convenience)
\begin{eqnarray}
L^{conf}(u_0,eB)       &=& 2 \int_{u_0}^\infty du  \left(\frac{R}{u}\right)^{3/2} f^{-1} \sqrt{\frac{u_0^8 f_0  A_0}{u^8 f A - u_0^8 f_0  A_0}} \nonumber \\
&\stackrel{\zeta = (u/u_0)^{-3}}{=}& \frac{2}{3} \frac{R^{3/2}}{\sqrt {u_0}} \sqrt{f_0 A_0} \int_0^1 d\zeta \frac{f^{-1} \zeta^{1/2}}{\sqrt{fA - f_0 A_0 \zeta^{8/3}}}.  \label{LconfifvB}
\end{eqnarray}

To model magnetic catalysis, we keep $L^{conf}(u_0,eB)$ fixed, for each flavour, at its flavour-independent value $L$ for  $eB=0$. From the viewpoint of the boundary field theory, the flavour branes are extensive objects that stretch out infinitely far, viz.~from $u=\infty$ to $u=u_0$ into the higher-dimensional bulk space. As such, from this asymptotic perspective, it would appear that it would cost an infinite amount of energy to move the branes at $u=\infty$.  Keeping $L$ fixed  thus seems to be a sensible boundary condition, one also used in e.g.~\cite{Peeters:2006iu,Preis:2010cq}, to probe the effects of the bulk dynamics in the presence of the external field.  The value of $L$, ranging from $\frac{\delta \tau}{2}$ to negligible w.r.t. $\frac{\delta \tau}{2}$, determines how much of the bulk dynamics is probed, ranging respectively from all to none.

\begin{figure}[h]
\centering
\subfigure[]{\includegraphics[scale=0.7]{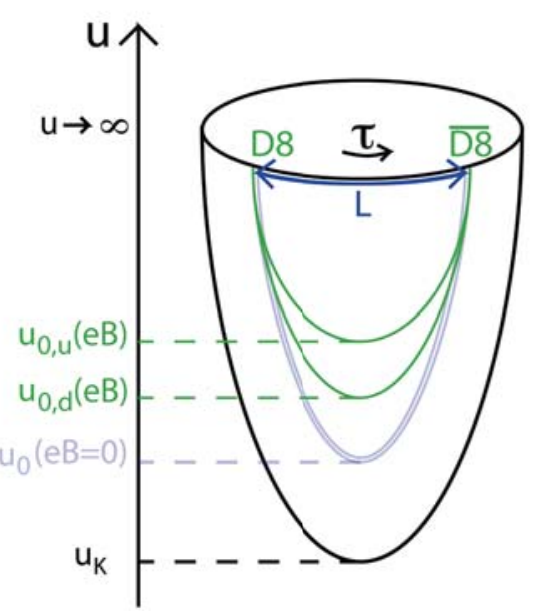} \label{embeddingBa}}
\hspace{4cm}\subfigure[]{\includegraphics[scale=0.7]{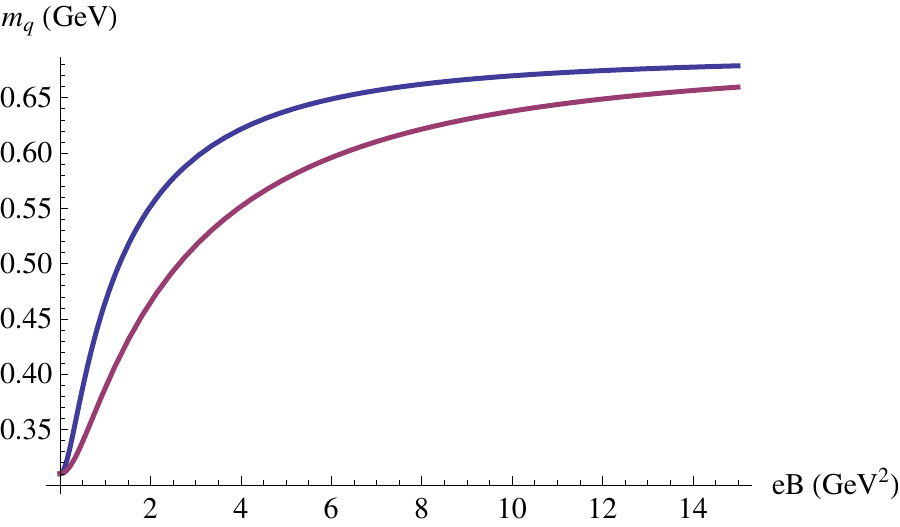} \label{embeddingBb} }
\caption[]{(a) Up and down flavour brane embedding in the presence of a magnetic field, (b) corresponding constituent quark masses $m_{q,u}$ (blue) and $m_{q,d}$ (red) for parameter choice (\ref{values}).}\label{embeddingB}
\end{figure}

So, as we keep $L^{conf}$ fixed to its value $L$ at zero magnetic field, from here on also explicitly writing the $M$-dependence of $L^{conf}$ through $R$ and $2\pi\alpha'$, we can extract the $eB$-dependence of $u_0$ and of the corresponding constituent quark masses $m_{q,u}$ and $m_{q,d}$, which correspond to the mass (\ref{mq}) of a string stretched between one of the flavour branes and the cut-off at $u_K=M^{-1}$ \cite{Aharony:2006da}:
\[
L^{conf}_\ell(u_0,eB,M) = L \quad \Longrightarrow \quad u_{0,\ell}(eB,L,M) \quad \Longrightarrow \quad m_{q,\ell}(eB,L,M).
\]
Using the correspondence between $L$ and $M$ as plotted in Figure~\ref{b} such that $m_\rho=0.776$ GeV, we obtain $m_q(eB,M)$ or $m_q(eB,L)$. This turns out to be a rising function of $eB$ for all choices of $M$ or $L$, modeling chiral magnetic catalysis, as already noted in \cite{Johnson:2008vna} for the single-flavour case.
The magnetic field thus indeed boosts the chiral symmetry breaking, reflected in a stronger bending of the flavour branes, and the breaking is stronger for the up- than for the down-flavour, see Figure \ref{embeddingBa}. This was expected, since the up-quark couples twice as strong to the magnetic field.
For the particular choice of parameters (\ref{values}), $m_q$ is plotted in Figure~\ref{embeddingBb}.
Both constituent masses show a quadratic dependence on $eB$ for small magnetic fields, which then becomes nearly linear, and eventually thrives to saturation at very large $eB$. This linear behaviour is in relative agreement with recent lattice estimates of the chiral condensate\footnote{The concept of the chiral condensate is not well-defined in the SSM, we therefore use the constituent quark mass as a measure for the chiral symmetry breaking.} in \cite{D'Elia:2011zu,Bali:2011qj,Ilgenfritz:2012fw}. A linear behaviour, inspired by chiral perturbation theory, was fitted to the SU(2) lattice data for the chiral condensate in \cite{Buividovich:2008wf}, while SU(3) data was fitted with a $(eB)^{3/2}$ behaviour in \cite{Braguta:2010ej}; both these studies used 1 quenched flavour.
Let us however point out that the plots of the lattice chiral condensates in all cited works do seem to display a nontrivial curvature for small values of $eB$.

The stability of the flavour branes' embedding as shown in Figure~\ref{embeddingBa} is discussed in \cite{Callebaut:2011ab}. At sufficiently low values of $eB$, it is stable, but at higher values of $eB$, we should not necessarily trust our findings, as an instability could occur in the $\rho$ meson sector at $eB\sim m_\rho^2\sim 30\, m_\pi^2\, (\approx 0.6\,\text{GeV}^2)$, driving a condensation \cite{Chernodub:2010qx}. Evidently, this would lead to a different underlying action, describing the theory in the condensed phase. If this happens, also other models should take it into account.

\FloatBarrier

\subsection{$eB$-dependent embedding in deconfinement phase}

Next, we turn to the finite temperature case. The background at $T<T_c$ is identical to the zero temperature background up to the period $\beta = T^{-1}$ of Euclidean time, so nothing changes as compared to the $T=0$ case. Things get more interesting once we enter the deconfinement region. We again have
an induced metric on each flavour brane,
\begin{eqnarray}
ds^2_{\text{D}8} &=& \left(\frac{u}{R}\right)^{3/2} (\hat f dt^2 + \delta_{ij}dx^i dx^j)
+ \left(\frac{R}{u}\right)^{3/2} u^2 d\Omega_4^2+\left(\frac{u}{R}\right)^{3/2} \left[ \frac{1}{\hat f} \left(\frac{u}{R}\right)^{-3} + \frac{1}{u'^2}\right] du^2\,,
\end{eqnarray}
with periodicity of the $t$-circle given by
\begin{equation}
\delta t = T^{-1}= \frac{4\pi}{3} R^{3/2} u_T^{-1/2},
\end{equation}
from which one 
can determine the action in the deconfined phase, completely analogous to the derivation of the action in the confined phase.
For temperatures $T< T_{\chi,\ell}$, the $\ell$-brane's embedding remains $\cup$-shaped, with action
\begin{eqnarray}
{S}_\ell^{T<T_{\chi,\ell}}
&=& c_0 u_{0,\ell}^{7/2} \int_{1}^\infty dy \hspace{1mm} y \sqrt{y^3 A_\ell} \sqrt{\frac{1}{1 - \frac{\hat f_{0,\ell} A_{0,\ell}}{\hat{f}_\ell(y) y^3 A_\ell}y^{-5}}},
\end{eqnarray}
where $c_0= - 2 T_8 \mathcal{V}_4 V_4 g_s^{-1} R^{3/2}$,
$y=u/u_{0,\ell}$, $y_{T,\ell}=u_T/u_{0,\ell}$, $\hat{f}_\ell(y) = 1-(y_{T,\ell}/y)^3$ and $\hat{f}_{0,\ell} = 1-y_{T,\ell}^3$. If $T>T_{\chi,\ell}$, the $\ell$-branes are falling straight down, $u' = \infty$, with action
\begin{eqnarray}
{S}_\ell^{T>T_{\chi,\ell}}
&=& c_0 u_{0,\ell}^{7/2} \int_{y_{T,\ell}}^\infty dy \hspace{1mm} y \sqrt{y^3 A_\ell}.
\end{eqnarray}
The chiral transition temperature $T_{\chi,\ell}$ is the temperature for which $\Delta S_{\ell}$ becomes zero \cite{Aharony:2006da},
with
\begin{eqnarray}\label{kritiek}
\Delta S(u_0,eB,y_T) &=& \text{action}_{\cup\text{-shape}} -  \text{action}_\text{straight}\,.
\end{eqnarray}
The correspondence between $u_{0}$ and $L$ in the deconfined phase is modified into 
 (again suppressing the flavour index here)
\begin{eqnarray}
L^{dec}(u_0,eB,y_T)  &=& \frac{2}{3} \frac{R^{3/2}}{\sqrt {u_0}} \sqrt{\hat f_0 A_0} \int_0^1 d\zeta \frac{\hat f^{-1/2} \zeta^{1/2}}{\sqrt{\hat fA - \hat f_0 A_0 \zeta^{8/3}}}. \label{LeindigeT}
\end{eqnarray}

\begin{figure}[h]
\centering
\subfigure[]{\includegraphics[scale=0.8]{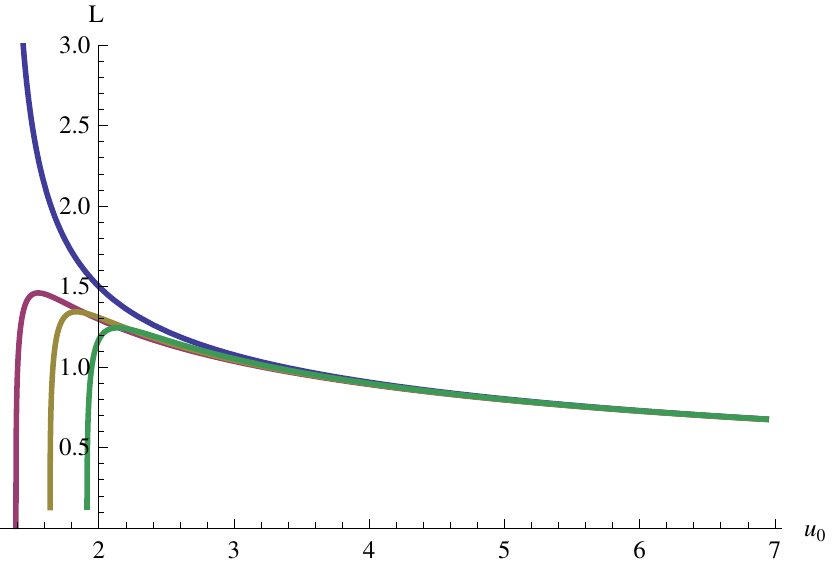} \label{Lu0fig}}
\hspace{1cm}
\subfigure[]{\includegraphics[scale=0.8]{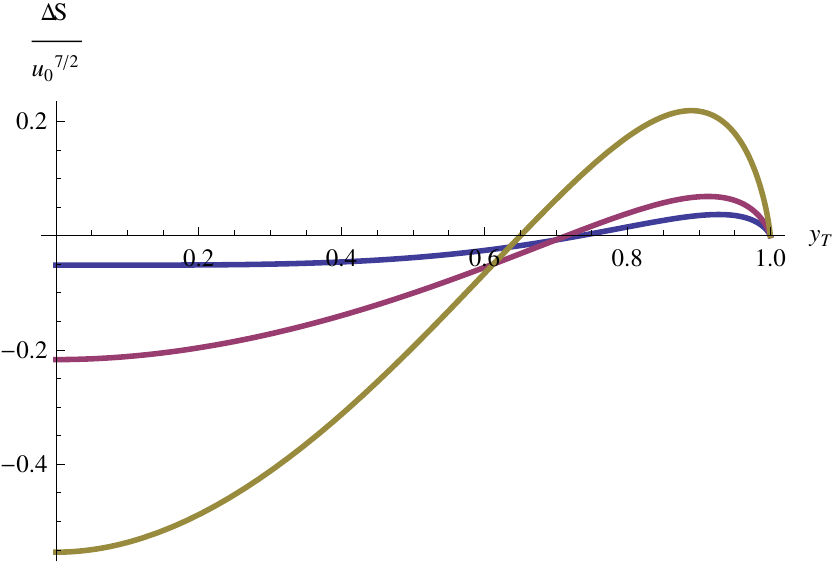}  \label{DeltaSyT} }
\caption[]{(a) $L^{conf}$ (blue) and $L^{dec}$ ($\text{GeV}^{-1}$) for $T=T_c$ (red)  and increasing values of $T(>T_c)$ at $eB=0$, (b) $\Delta S/u_0^{7/2}$ as a function of $y_T$ for $eB=0$ (blue), 0.5 (purple) and 1.6 GeV$^2$ (yellow). Both figures for $M=0.7209$ GeV.  }\label{}
\end{figure}

As before, we will hold the asymptotic separation fixed at its starting value $L$ at $eB=0$ and $T=0$.
This allows us to determine the $eB$- and $T$-dependence of $u_0$ (from here on also explicitly writing the dependence on $M$, through $R$ and $2\pi\alpha'$):
\begin{equation} \label{u0BTL}
L^{dec}(u_0,eB,y_T,M) = L \quad \Longrightarrow \quad u_0(eB,y_T,L,M).
\end{equation}
From Figure~\ref{Lu0fig} it can be seen that the one-to-one correspondence between $u_0$ and $L^{conf}$ is not preserved in the deconfinement phase, where each value of $L^{dec}$ corresponds to two possible values of $u_0$, as long as it does not exceed its maximum possible value (i.e.~as long as $T<T_\chi$). We numerically verified that the energetically favoured solution for $u_0$ is the largest one, consistent with the intuition that the lower-$u_0$ solution contains more energy as it probes a larger portion of the background. Keeping $L$ fixed during the deconfinement transition causes a jump in $u_0$, as well as in the constituent quark and meson masses \cite{Peeters:2006iu}.

With the expression found for $u_0$, the expression for $\Delta S$ at fixed $L$ is also known,
\begin{equation}
\Delta S(u_0(eB,y_T,L,M),eB,y_T,M) \equiv \Delta S(eB,y_T,L,M),
\end{equation}
so the chiral temperature can be determined from the point where the $\cup$-shaped embedding breaks into separated branes, i.e. when $\Delta S = 0$ (see Figure~\ref{DeltaSyT}):
\[
\Delta S(eB,y_T,L,M) = 0 \quad \Longrightarrow \quad y_T^\chi(eB,L,M).
\]
The corresponding value of $u_0$ at the chiral transition is then given by
\[
u_0^\chi = u_0(eB,y_T^\chi(eB,L,M),L,M) \equiv u_0^\chi(eB,L,M).
\]
Plugging the obtained $y_T^\chi(eB,L,M)$ and $u_0^\chi(eB,L,M)$ into the definition for the chiral temperature
\begin{align*}
T_\chi  &= \frac{3}{4\pi} \sqrt{\frac{u_T^\chi}{R^3}} = \frac{3}{4\pi} \sqrt{y_T^\chi} \sqrt{\frac{u_0^\chi}{R^3}} = \frac{3}{4\pi} \sqrt{y_T^\chi(eB,L,M)} \sqrt{\frac{u_0^\chi(eB,L,M)}{R^3}} \equiv T_\chi(eB,L,M),
\end{align*}
we obtain $T_\chi(eB,L,M)$.

\begin{figure}[h]
  \centering
  \scalebox{0.7}{
  \includegraphics{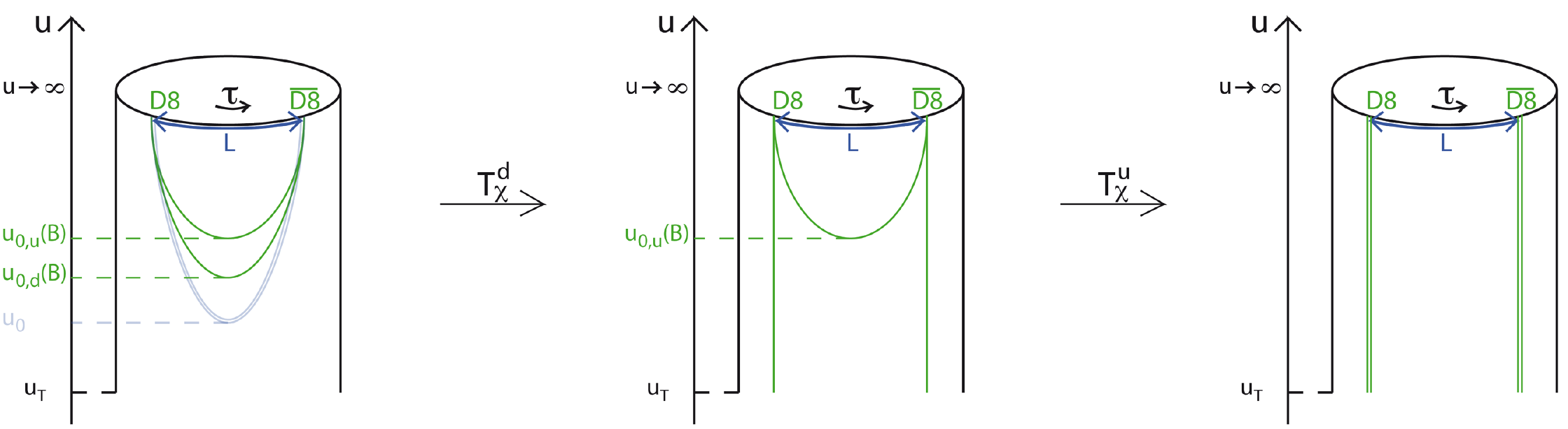}}
  \caption{Embeddings in the deconfined phase with magnetic field. }\label{TchiralSS}
\end{figure}

From the parameter discussion at $eB=0$ we know the value of the fixed asymptotic separation given a value for $M$ such that $m_\rho=0.776$ GeV (Figure~\ref{b}), hence we obtain $T_\chi(eB,M)$ or $T_\chi(eB,L)$, to be compared with the deconfinement temperature $T_c = M/(2\pi)$.
The deconfinement temperature $T_c$ will not change as it is determined from the background D4-brane metric, which could only become $eB$-dependent when the backreaction of the D8-branes would be taken into account.  Or, field theoretically: $T_c$ is $eB$-independent in a quenched setup, because the magnetic field can only couple to the neutral gluons indirectly via the quark interactions.
For every choice of $M$ (or $L$), $T_\chi(eB)$ rises with $eB$ (``chiral magnetic catalysis'').
But depending on the choice, there will or will not arise a split between $T_\chi$ and $T_c$.
Doing this for each flavour, we find $T_\chi^u(eB,L)$ and $T_\chi^d(eB,L)$, with $T_\chi^u$ consistently higher than $T_\chi^d$ for a given value of $L$, as expected. This leads to an intermediate phase where the chiral symmetry for up quarks is still broken while the chiral symmetry for down quarks is already restored:
\[
U(1)_V^u \times U(1)_V^d \stackrel{T_\chi^d}{\rightarrow}  U(1)_V^u \times  (U(1)_L \times U(1)_R)^d \stackrel{T_\chi^u}{\rightarrow}(U(1)_L \times U(1)_R)^u \times (U(1)_L \times U(1)_R)^d,
\]
as sketched in Figure~\ref{TchiralSS}.

\begin{figure}[h]
\centering
\subfigure[]{\includegraphics[scale=0.59]{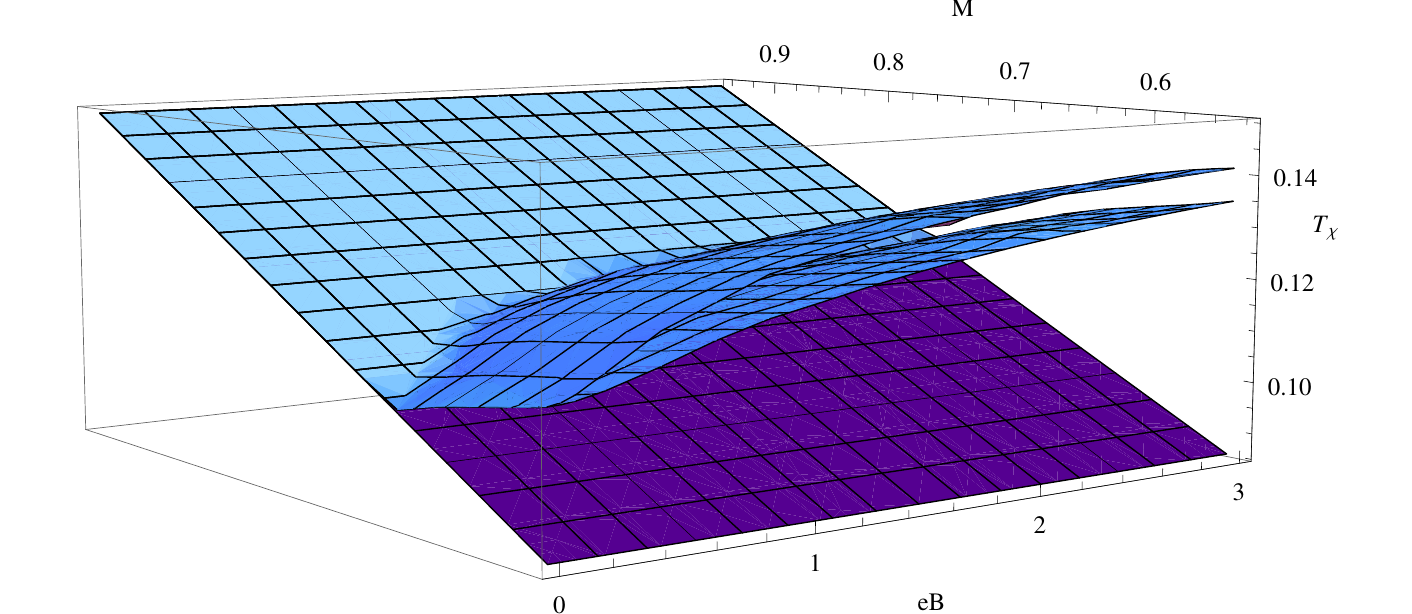} \label{TchiBM}}
%\hspace{4cm}
\subfigure[]{\includegraphics[scale=0.75]{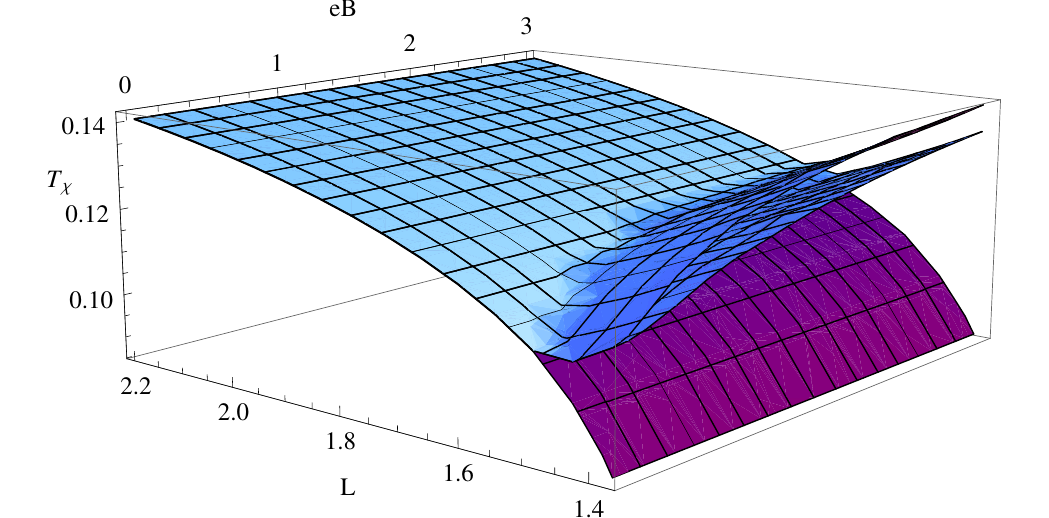} \label{TchiBL}}
\caption[]{(a) $T_\chi^u$ (GeV) (upper blue surface) and $T_\chi^d$ (GeV) (lower blue surface) as functions of $eB$ ($\text{GeV}^2$) and $M$ (GeV) compared to $T_c(M)$ (GeV) (purple), (b) same with $M$-dependence replaced by $L$-dependence compatible with $m_\rho=0.776$ GeV. }\label{Tchifig}
\end{figure}

\begin{figure}[h]
\centering
\subfigure[]{\includegraphics[scale=0.65]{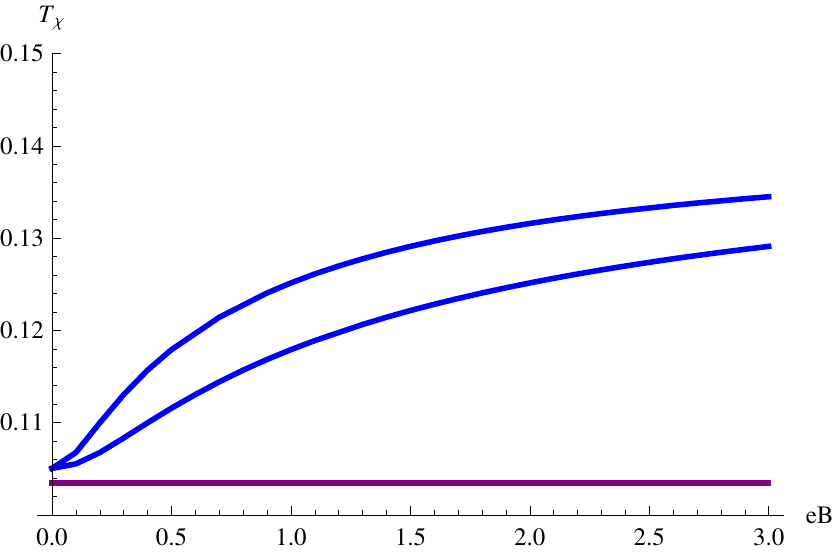} \label{T65}}
%\hspace{4cm}
\subfigure[]{\includegraphics[scale=0.65]{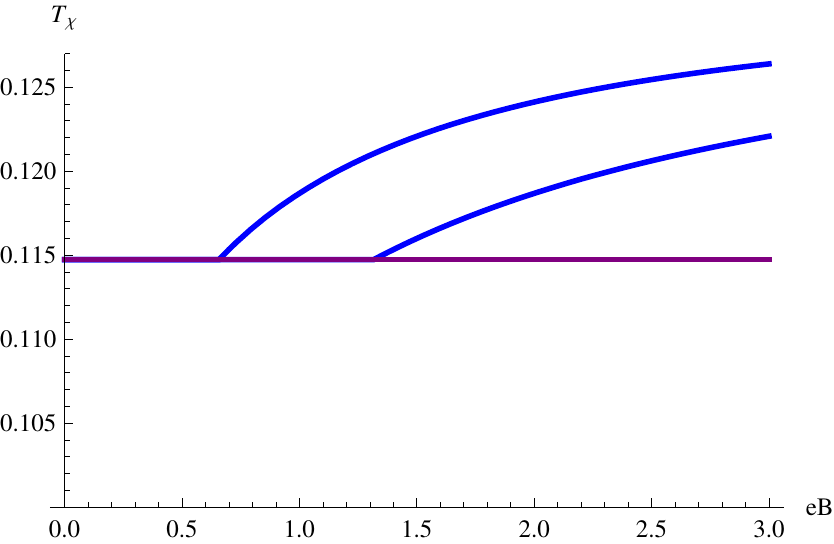} \label{T7209}}
\subfigure[]{\includegraphics[scale=0.65]{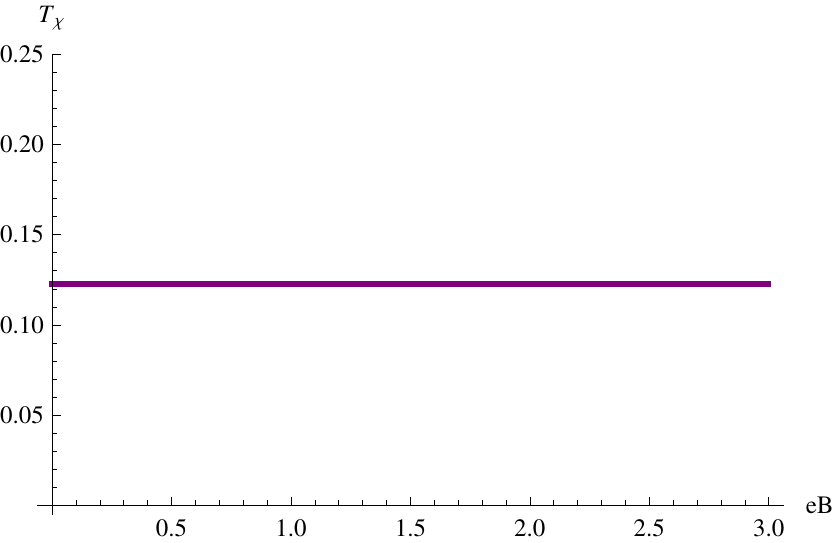}  \label{T77}}
\caption[]{Cross sections of Figure~\ref{Tchifig} for (a) $M=0.65$ GeV, (b) $M=0.7209$ GeV and (c) $M=0.77$ GeV, respectively corresponding to $m_q(eB=0) = 0.357, 0.310$ and $0.272$ GeV and $T_c=0.103, 0.115$ and 0.123 GeV. The appearance of a split between $T_\chi$ (GeV) (blue) and $T_c$ (GeV) (purple) depends on the choice of $M$, or equivalently $L$. }\label{Tchicrosssections}
\end{figure}

\begin{figure}[h]
  \centering
  \scalebox{0.6}{
  \includegraphics{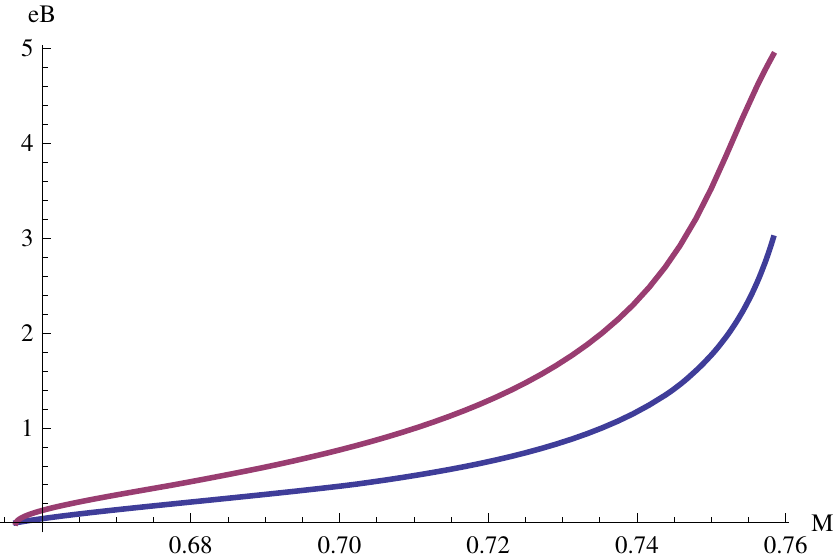}}
  \caption{Value of $eB$ where $T_\chi^u$ (GeV) (blue) resp.~$T_\chi^d$ (GeV) (purple) becomes larger than $T_c$ (GeV) for confinement scale values 0.657 GeV $< M < 0.767$ GeV.} \label{BcM}
\end{figure}

In Figure~\ref{Tchifig} the $(T,M,eB)$ and $(T,L,eB)$ phase diagrams of the two-flavour non-antipodal SSM are plotted. This generalizes the $N_f=1$ SSM phase diagram  in Figure 7 of \cite{Aharony:2006da} to the $N_f=2$ magnetic case. For set-ups with large values of $M$, namely $M>0.767$ GeV corresponding to $m_q(eB=0) < 0.274$ GeV or $L>1.681$ GeV$^{-1}$,
there is no split between $T_\chi(eB,M)$ and $T_c=M/(2\pi)$, no matter how large the applied magnetic field is. This is a consequence of the saturation of the rising of $T_\chi$ with $eB$.
In a SSM with $M<0.657$ GeV, corresponding to $m_q(eB=0) > 0.353$ GeV or $L<1.473$ GeV$^{-1}$, there is already a split between chiral and deconfinement transition before the magnetic field is turned on: $T_\chi(eB=0,M) > T_c$, which becomes larger as $eB$ increases. This regime is probably the least physically relevant, as the values for constituent quark masses are too large and the values for the deconfinement temperature smaller than 0.105 GeV, which is rather small compared to the chiral limit value we can extrapolate ``by hand'' from \cite{Bornyakov:2009qh}, giving $T_c\sim 0.150~\text{GeV}$. The third possible case is that the value of $M$ is such, 0.657 GeV $< M < 0.767$ GeV ($\sim 0.274$ GeV $< m_q(eB=0) < 0.353$ GeV or 1.473 GeV$^{-1} < L < 1.681$ GeV$^{-1}$), that $T_\chi(eB=0,M) = T_c$ but a split between $T_\chi$ and $T_c$ arises at some value of $eB$, plotted in Figure~\ref{BcM}.
For each of the above possible cases, an exemplary cross section of the $(T,M,eB)$ phase diagram is shown in Figure~\ref{Tchicrosssections}, the middle one corresponding to the best matching parameters for reproducing a reasonable $m_q(eB=0)\approx 0.310$ GeV, although the corresponding value for $T_c \approx 0.115$ GeV is still on the small side.
A deconfinement temperature $T_c \approx 0.150$ GeV would correspond to $M=0.942$ GeV, very close to the antipodal value so in the regime where no split arises between $T_\chi$ and $T_c$, leading however to an unphysically\footnote{This might be related to the shortcoming of the SSM (in the used form, not considering possible modifications as in \cite{Dhar}) that the bare quark masses always remain zero.} small $m_q(eB=0) \approx 0.046$ GeV. 

The papers \cite{Johnson:2008vna,Preis:2010cq} also pointed out the splitting of the critical temperatures for the one flavour version of the SSM, but leaving the parameters of the SSM undetermined, making an explicit comparison with other approaches harsh. The explicit breaking of the global flavour symmetry by the different electromagnetic coupling of the up and down flavour is also now taken into account for the first time, leading to a split between the two chiral transitions themselves.

The fact that a split between $T_\chi$ and $T_c$ can emerge only for sufficiently small values of the asymptotic brane separation $L$, i.e. sufficiently close to an NJL effective description of QCD, seems to be supported by NJL model calculations \cite{Gatto:2010pt}, see also the discussions in \cite{Mizher:2010zb,Boomsma:2009yk,Klevansky:1992qe}, that seemingly contrast with lattice data.
Selecting the holographic parameters in a way that brings the SSM as close as possible to (the chiral limit of) QCD, rather leads to a picture of the form of 
Figure~\ref{T7209} or \ref{T77}, namely no split at all or a small split that only emerges at rather large values of $eB$. Our findings are in this perspective consistent with lattice data of \cite{Ilgenfritz:2012fw}, where a split was neither reported. However, we must also repeat here that our results are obtained in a quenched framework, hence important QCD effects at the level of transitions can be missing (e.g.~pion loop effects). In particular, magnetic effects on the deconfinement temperature cannot be taken into account in the SSM without including backreaction of the probe branes on the background D4 metric.

\subsubsection{Remark on the antipodal SSM}
In the original antipodal Sakai-Sugimoto model, with $u_0=u_K$ and the asymptotic separation $L$ taking
its maximum possible value, the embedding of the flavour
branes is unaffected by the presence of the magnetic field.
From this we can conclude that the antipodal Sakai-Sugimoto
model is unable to capture the magnetically induced explicit
breaking of chiral symmetry, as well as the chiral magnetic
catalysis. Chiral symmetry restoration and deconfinement coincide
for all values of the magnetic field.

\section{Summary}

In conclusion, we have investigated the phase diagram of the two flavour version of the non-antipodal Sakai-Sugimoto model in the presence of a temperature $T$ and external magnetic field $eB$.
In particular we payed attention to fixing the holographic parameters, presenting a discussion of how they can be fixed by matching to carefully chosen QCD input parameters, in order to be able to present the phase diagram and related results in physical GeV units. This makes comparison to other approaches more direct. We indeed could compare our results with lattice and NJL results, the main conclusion being that the SSM results are consistent with other quenched settings that are able to model chiral magnetic catalysis.

The main results are presented in the $(T,L,eB)$ phase diagram in Figure~\ref{Tchifig} and cross sections of that plot for different values of $L$ in Figure~\ref{Tchicrosssections}. Here, $L$ is the asymptotic separation between the flavour probe branes.  Keeping $L$ fixed serves as a boundary condition for the bulk dynamics, the effective boundary theory ranging from the NJL-type for small $L$ to a chiral QCD-like theory where gluon dynamics are fully taken into account for maximal $L$. The value of $L$, i.e. the choice of the type of boundary model in a sense, determines if a split between the chiral and deconfinement temperature may or may not arise, as summarized in Figure~\ref{Tchicrosssections}. Due to the different coupling to the magnetic background of the $u$ and $d$ flavour brane, we also find a split between the separate chiral transition temperatures.

It remains a challenge to construct a holographic dual of realistic QCD that could also describe the complicated finite temperature (above and below $T_c$) behaviour of the chiral condensate as found in the latest lattice results  \cite{Bali:2011qj}. This would require taking backreaction of the flavour branes on the background metric into account.
A holographic model that could be interesting to look at from this point of view, is the Kuperstein-Sonnenschein model \cite{Kuperstein:2008cq}, in which chiral magnetic catalysis was observed \cite{Alam:2012fw} in the non-backreacted case, and which was extended to include backreaction (be it perturbatively and for vanishing magnetic field) in \cite{Ihl:2012bm}, using a smearing technique. A downside of this model is however that it does not incorporate confinement due to the choice of the background in which the flavour branes are placed.

We end by noticing that the magnetic phase diagram could become even more intricate, as demonstrated in \cite{Chernodub:2010qx}: the QCD vacuum could become superconducting at sufficiently strong magnetic field. This matter is currently under investigation also in the holographic framework, see also \cite{Callebaut:2011ab,ammon}. We hope the presented results will stimulate further research in the area of QCD in a strong magnetic background, where many exciting new physics could be awaiting discovery.  The gauge-gravity correspondence can offer viable input, next to QCD model and lattice based approaches.

\section*{Acknowledgments}
We thank the Research-Foundation Flanders (FWO-Vlaanderen) for financial support. We are grateful to D.~Boer, E.~Caceres, M.~N.~Chernodub, E.~S.~Fraga, A.~J.~Mizher, L.~F.~Palhares, A.~Rebhan, A.~Schmitt and H.~Verschelde for insightful discussions.

\end{document}